# Wetting-phase relative permeability in porous media with bi-modal pore size distributions

Behzad Ghanbarian

Porous Media Research Lab, Department of Geology, Kansas State University, Manhattan KS 66506 USA (Email address: ghanbarian@ksu.edu)

**Abstract**

Modeling fluid flow in dual-porosity media with bi-modal pore size distributions has practical applications to understanding transport in multi-scale systems such as natural soils. Dual-porosity media are typically formed of two domains: (1) structure and (2) texture. The former mainly incorporates macropores, while the latter contains micropores. Although there exist models based on the series-parallel tubes approach, here we apply concepts from critical path analysis, a theoretical technique from statistical physics, to estimate water relative permeability ($k_{rw}$) in dual-porosity media. For this purpose, we use two datasets from the literature collected under two different cultivation conditions: (i) conventional tillage (CT) and (ii) non-tillage (NT). Each dataset consists of 13 soil samples for which capillary pressure curve and water relative permeability were measured at 500 data point over a wide range of water saturation. We estimate the water relative permeability from the measured capillary pressure curve using two methods: (1) critical path analysis (CPA), and (2) series-parallel tubes (vG-M), both models adapted for dual-porosity media. Comparing the theoretical estimations with the experimental measurements shows that CPA resulted in more accurate $k_{rw}$ estimations than vG-M. We demonstrate that precise estimation of $k_{rw}$ via CPA requires accurate characterization of capillary pressure curve and precise determination of the crossover point separating the structure domain from the texture one.

**1. Introduction**

Multi-scale porous media with multi-modal pore size distributions have been frequently observed in the nature (Dal Ferro et al., 2013; Dörner et al., 2010; Farahani et al., 2019; Millán and González-Posada, 2005). Such systems, however, are more heterogeneous in terms of pore sizes than media with uni-modal pore size distributions. Accordingly, modeling fluid flow and transport in multi-scale media is more complicated and requires incorporating various types of complexities.

In the literature, different dual-porosity models based on a "bundle of capillary tubes" concept were proposed to estimate water relative permeability from capillary pressure curve (Durner, 1994; Kutílek et al., 2009; Mohanty et al., 1997; Ross and Smettem, 1993). Such a concept, despite its widespread use, is a severely distorted idealization of natural porous media. The parallel (Purcell, 1949) and series-parallel (Mualem, 1976) tubes models are illustrated in Fig. 1. In both approaches there is no connection between the tubes. In addition, individual pores span the entire sample, regardless of its size. In contrast to the parallel tubes approach in which pores have a constant cross-sectional area along the tube, the series-parallel tubes approach considers connected tubes in series to address changes in cross-section (Fig. 1).

Othmer et al. (1991) are among the first who combined the van Genuchten capillary pressure curve model (van Genuchten, 1980) with Mualem's approach, both adapted for dual-porosity media. By comparison with experiments measured on Neuenkirchen loam samples taken from two depths (i.e., 15 and 60 cm), those authors found more reasonable estimations of $k_{rw}$ than the uni-modal model of van Genuchten (1980). Othmer et al. (1991) defined the effective water saturation and water relative permeability simply as $(S_{w1} + S_{w2})$ and $(k_{rw1} + k_{rw2})$, respectively (see their Eqs. 7 and 8). If weighting factors corresponding to the fraction of each mode are incorporated, one has (Priesack and Durner, 2006)

$$S_e = \frac{S_w - S_{wr}}{1 - S_{wr}} = \sum_{i=1}^{2} w_i [1 + (\alpha_i P_c)^{n_i}]^{-m_i} \tag{1}$$

$$k_{rw} = (\sum_{i=1}^{2} w_i S_w)^{0.5} \left[ \frac{\sum_{i=1}^{2} w_i \alpha_i \left[1 - \left(1 - S_w^{1/m_i}\right)^{m_i}\right]}{\sum_{i=1}^{2} w_i \alpha_i} \right]^2 \tag{2}$$

where $S_e$ is the effective saturation, $S_w$ is the water saturation, $S_{wr}$ is the residual water saturation, $w$ represents the weighting factor, $P_c$ is the capillary pressure, and $\alpha$, $n$ and $m$ denote van Genuchten capillary pressure curve model parameters.

In another study, Zhang and van Genuchten (1994) proposed closed-form equations to estimate $k_{rw}$ from the bi-modal capillary pressure curve. Using more than 20 soil samples, they reported good agreement between estimated water relative permeabilities and measured values. They also found more satisfactory estimations in comparison with the uni-modal model of van Genuchten-Mualem (Mualem, 1976; van Genuchten, 1980).

Coppola (2000) investigated the capability of bi-modal models for describing capillary pressure curve and estimating water relative permeability of 18 aggregated soil samples. He showed that the uni-modal model of van Genuchten (1980) was unable to accurately characterize measured capillary pressure curves, particularly near the saturation point. Coppola (2000) showed that water relative permeability was more precisely estimated, within one order of magnitude of measurements, when bimodal models were used.

By combining the multi-modal representation of the van Genuchetn capillary pressure curve with the generalized form of the Mualem model, Priesack and Durner (2006) derived a simple analytical expression to estimate water relative permebaility in soils with multi-modal pore size distributions. Three samples (including an aggregated loam soil, a sandy loam soil, and macroporous silt loam) were analyzed to demonstrate the applicability and flexibiltiy of bi-modal and tri-modal models and predictibility of water relative permeability. Although they showed the usefulness of the multi-modal representations of both capillary pressure curve and water relative permebaility, their model evaluation was limited to three soil samples.

Recently, Reynolds (2017) segregated soil pore space into three domains: (1) structure, (2) matrix, and (3) residual. The van Genuchten model was accordingly modified to reflect the three domains. He defined separate soil physical quality (SPQ) indicators for water and air storage in the pore space such as porosity, available water capacity, pore size distribution, as well as tension heads and water contents for air-entry and field capacity. Reynolds (2017) fit his model to measured capillary pressure curve and water relative permeability of three samples i.e., repacked diatomite pellets, repacked soil aggregates, and intact soil with biopores and cracks. He stated that, "Although overall SPQ for the bulk medium may be good, structure and matrix SPQ may be limited to poor, with structure tending to be water-limited and potentially prone to leaching losses, and matrix tending to be poorly aerated and potentially prone to greenhouse gas generation."

Concepts of critical path analysis (CPA) from percolation theory have been successfully used to model hydraulic properties of porous media (Hunt et al., 2014; Hunt, 2001; Hunt and Sahimi, 2017). Based on CPA transport in a network of pores is dominated by those pores whose sizes are greater than some critical size. Accordingly, pores smaller than the critical pore size make trivial contribution to the overall transport. Imagine a pore network constructed of pores of various sizes shown in Fig. 2a. Following Friedman and Seaton (1998), let us remove all pores from the network. We then replace them in their original locations in a decreasing order from the largest to the smallest pore size. As the first largest pores are replaced, there is still no percolating cluster (Fig. 2b). However, after a sufficiently large fraction of pores are replaced within the network, a sample-spanning cluster forms and the system starts percolating (Fig. 2c).

Ghanbarian-Alavijeh and Hunt (2012) combined the pore-solid fractal (PSF) model of Perrier et al. (1999) with the critical path analysis approach and developed a new model to estimate $k_{rw}$ in porous media with uni-modal pore size distributions. Comparing with more than 100 soil samples from the UNSODA database indicated that their proposed model resulted in more accurate estimates of $k_{rw}$ than the van Genuchten-Mualem model. However, years later, Ghanbarian et al. (2016) and Ghanbarian and Hunt (2017) showed that critical path analysis estimations of $k_{rw}$ could be even improved by incorporating the effect of pore-solid interface roughness.

Hunt et al. (2013) extended the uni-modal model of Ghanbarian-Alavijeh and Hunt (2012) to media with bi-modal pore size distributions (two fractal regimes). They adapted the PSF model for dual-porosity media and presented the following capillary pressure curve model

$$S_w = \begin{cases} 1, & P_c < P_d \\ 1 - \frac{\beta_1}{\phi}\left[1 - \left(\frac{P_c}{P_d}\right)^{D_1-3}\right], & P_d < P_c < P_x \\ \frac{\phi_2}{\phi} - \frac{\beta_2}{\phi}\left[1 - \left(\frac{P_c}{P_x}\right)^{D_2-3}\right], & P_c > P_x \end{cases} \quad (3)$$

in which $D_1$ and $D_2$ are the fractal dimensions, $\phi_1$ and $\phi_2$ are the porosities, and $\beta_1$ and $\beta_2$ are PSF model parameters of the first and second modes, respectively, $\phi$ is the total porosity, $P_d$ is the displacement capillary pressure (or the air entry pressure), and $P_x$ is the capillary pressure at the crossover point.

By combining Eq. (3) with the critical path analysis approach, Hunt et al. (2013) proposed the following bi-modal model for $k_{rw}$:

$$k_{rw} = \begin{cases} \left[\frac{\beta_1 - \phi + \phi S_w}{\beta_1}\right]^{\frac{3}{3-D_1}}, & S_w > \phi_2/\phi \\ \left(\frac{\beta_1 - \phi_1}{\beta_1}\right)^{\frac{3}{3-D_1}} \left[\frac{\beta_2 - \phi_2 + \phi S_w}{\beta_2}\right]^{\frac{3}{3-D_2}}, & S_w < \phi_2/\phi \end{cases} \quad (4)$$

Hunt et al. (2013) set $\beta_1 = \beta_2 = 1$, determined $D_1$, $D_2$, $\phi_1$ and $\phi_2$ from capillary pressure curve and estimated $k_{rw}$ for 8 soil samples selected from the UNSODA database. They found that $k_{rw}$ estimations from the capillary pressure curve agreed relatively well with the measured one only at high water saturations, while their model underestimated water relative permeability at intermediate and low saturations.

The power-law capillary pressure curve model proposed by Hunt et al. (2013), Eq. (3), assumes that the displacement capillary (or air entry) pressure in the second mode is the same as the crossover pressure ($P_x$). However, since the power-law model is discontinuous in form, the two quantities are not necessarily the same. Consequently, one needs to revise Eq. (3) to better represent experimental measurements. We show that such a modification results into better agreement between $k_{rw}$ estimations and measurements. We should point out that most models developed in the literature were based on the bi-modal model of van Genuchten-Mualem and evaluated using only a few samples with limited number of measured data points, particularly on the capillary pressure curve. The latter, the restriction on measured points, results into uncertainties in the water relative permeability estimation. In addition, the reliability of the critical path analysis approach in the estimation of $k_{rw}$ has not yet been compared to that of the series-parallel tubes method in dual-porosity media. Thus, the main objectives of this study are to: (1) apply concepts from critical path analysis to dual-porosity media by revisiting the Hunt et al. (2013) model, (2) evaluate $k_{rw}$ estimation from accurately characterized $S_w$-$P_c$ curve using a database with 26 soil samples, and (3) compare the accuracy of the bi-modal CPA model with that of the bi-modal vG-M model.

## 2. Materials and Methods

The database used in this study is from Schwen et al. (2015) and includes 26 soils. Samples were collected from an arable field in Hollabrunn, Austria, under two different soil cultivation conditions: (1) conventional tillage with moldboard plowing and seedbed preparation by a harrow (hereafter CT), and (2) non-tillage with direct seeding (hereafter NT).

To measure bulk density, samples were oven dried at 105 °$C$ for 24 hours and accordingly total porosity was determined by assuming a particle density of 2.65 g cm$^{-3}$. Saturated volumetric water content ($\theta_s$), the amount of water required to occupy all the pores, was measured by gradually saturating samples from the bottom using deaerated water. It is worth pointing out that although we reported the total porosity value for each sample in Table 1, we used $\theta_s$ instead of $\phi$ for further analyses and estimating $k_{rw}$ from the capillary pressure curve meaning that $\theta_s = \phi_1 + \phi_2$. Recall that $\phi_1$ and $\phi_2$ are the porosities corresponding to the structure and texture domains, respectively. The $S_w$-$P_c$ and $S_w$-$k_{rw}$ curves, each of which includes 500 measured data points, were determined using the evaporation method with extended range of measurements. The dewpoint potentiometry method was used for measurements between 0.3 and 72.4 MPa pressures. Further detail about measurements can be found in Schwen et al. (2015, 2014).

Figure 3a shows the measured capillary pressure ($S_w$-$P_c$) curve for sample CT_1. To calculate $P_x$, we plotted $\Delta S_w / \Delta \ln(P_c)$ against $P_c$ and found the local minimum between the two modes (see Fig. 3b). The value of $\phi_2$ was then computed from $\theta_s S_{wx}$ in which $S_{wx}$ represents the water saturation corresponding to the crossover pressure $P_x$. We then determined $\phi_1$ from $\theta_s - \phi_2$.

As can be observed in Fig. 3b, the derived pore size distribution follows the bimodal log-normal probability density function. In contrast to the van Genuchten (1980) model, which describes the capillary pressure curve (or pore size distribution) smoothly and continuously, the power-law model is truncated and discontinuous at the displacement (or air entry) pressure. In other words, in uni-modal pore size distributions, the power-law probability density function only represents pore space characteristics corresponding to smaller pore sizes (or higher capillary pressures). Consequently, to determine the Hunt et al. (2013) $k_{rw}$ model parameters including $D_1, D_2, \beta_1, \beta_2, \phi$ and $\phi_2$, in this study we revisited the capillary pressure curve model and modified Eq. (3) as follows

$$S_w = \begin{cases} 1, & P_c < P_{d1} \\ 1 - \frac{\beta_1}{\phi}\left[1 - \left(\frac{P_c}{P_{d1}}\right)^{D_1-3}\right], & P_{d1} < P_c < P_x \\ S_{wx}, & P_x < P_c < P_{d2} \\ \frac{\phi_2}{\phi} - \frac{\beta_2}{\phi}\left[1 - \left(\frac{P_c}{P_{d2}}\right)^{D_2-3}\right], & P_c > P_{d2} \end{cases} \quad (5)$$

In contrast to Eq. (3), Eq. (5) assumes that the displacement pressure in the second mode $P_{d2}$ is different from the crossover pressure $P_x$.

Equation (5) with $\phi$ replaced with $\theta_s$ was fit to the capillary pressure curve using the CurveFitting toolbox of MATLAB. For this purpose, the two top and the two bottom functions in Eq. (5) were independently fit to the first (structure) and second (texture) domains, respectively, and parameters $D_1$, $D_2$, $\beta_1$, $\beta_2$, $P_{d1}$, and $P_{d2}$ were optimized. Figures 3c and 3d show Eq. (5) fitted to the measured data for sample CT_1. The optimized values of parameters in Eq. (5) are presented in Table 1 for each sample. Table 1 also summarizes the optimized parameters of the bimodal van Genuchten capillary pressure curve, Eq. (1), reported by Schwen et al. (2015); see their Table 1.

To compare statistically the accuracy of vG-M and CPA models, respectively Eqs. (2) and (4), in the estimation of $k_{rw}$, the root mean square log-transformed error (RMSLE) parameter was determined as follows

$$RMSLE = \sqrt{\frac{1}{N}\sum_{i=1}^{N}[\log(k_{rw})_{est} - \log(k_{rw})_{meas}]^2} \quad (6)$$

where $N$ is the number of data points, and indices *est* and *meas* represent estimated and measured, respectively.

## 3. Results and Discussion

### 3.1. Estimating bi-modal fractal capillary pressure curve model parameters

Table 1 lists the value of porosity $\phi$ and saturated volumetric water content $\theta_s$ for each sample. We found, on average, $\theta_s = 0.92\phi$ and the constant 0.92 not greatly different from 0.93 and 0.90 reported by Williams et al. (1992) and Pachepsky et al. (1999), respectively. The difference between $\phi$ and $\theta_s$ values is typically attributed to trapped air bubbles in the pore space during the saturation process. The average $\phi$ values for the CT and NT samples are respectively 0.54 and 0.44, which indicates soil compaction under non-tillage conditions. The obtained results are in accord with those reported by Gómez et al. (1999), Lipiec et al. (2006), Soracco et al. (2019) and others. For example, Gómez et al. (1999) indicated that after 15 years their NT treatment achieved greater bulk density than the CT one.

Table 1 also presents the fitted values for the bimodal fractal capillary pressure curve model, Eq. (5), parameters for CT and NT samples. The effect of tillage on fractal behavior of soils has been an active research area (Eghball et al., 1993; Giménez et al., 1998, 1997; Perfect and Kay, 1995; Torre et al., 2018; Vázquez et al., 2006). We found that the average $D_1$ values, corresponding to the structure domain, were 2.582 and 2.851 for CT and NT samples, respectively. The greater the fractal dimension, the more complex the pore space (Ghanbarian and Hunt, 2017). Accordingly, the soil structure in NT samples is more heterogeneous than that in CT samples. This could be due to dynamic effects of wetting and drying, micro-organism activities, etc. under no-tillage circumstances. Generally speaking, we found $D_1 < D_2$ for CT samples, while $D_1 > D_2$ for NT ones, which clearly indicates the non-trivial effect of tillage on the pore space. Although Hunt et al. (2013)

reported $D_1 > D_2$ for seven out of eight samples selected from the UNSODA database (see their Table 3), Russell (2010) analyzed four samples and found $D_1 < D_2$ for half of them (see his Table 2).

For all 26 samples we found $P_{d2} > P_x$ implying that the truncated power-law probability density function does not characterize the entire range of pore sizes in the second mode, similar to the $P_c < P_{d1}$ range in the first mode. As can be seen in Eq. (5), for both $P_c < P_{d1}$ and $P_x < P_c < P_{d2}$ ranges water saturation is constant and does not vary with pressure. Our results show that $P_x$ and $P_{d2}$ values are not necessarily the same and forcing $P_{d2}$ to be equal to $P_x$ may result into inaccurate determination of $D$ and $\beta$ values and consequently imprecise estimation of $k_{rw}$.

In Eq. (5), the effect of residual water saturation ($S_{wr}$) is not explicitly incorporated. Using a terminology based on fractals and power-law pore size distribution, Ghanbarian et al. (2017) recently derived $\beta = \frac{\phi(1-S_{wr})r_{max}^{3-D}}{r_{max}^{3-D}-r_{min}^{3-D}}$. Based on their theoretic derivation, when $r_{min} \to 0$, $\beta$ can be even less than $\phi$ depending on the $S_{wr}$ value. Accordingly, both $\beta_1 < \theta_s$ and $\beta_2 < \phi_2$, reported in Table 1 for several samples, are theoretically supported.

### 3.2. Estimating water relative permeability

In this section, we first evaluate Eqs. (3) and (5) in the estimation of water relative permeability from the capillary pressure curve. Then, we compare the bi-modal CPA approach, Eq. (4), with the bi-modal vG-M model, Eq. (2), using the 26 soil samples studied here.

Figure 4 shows Eqs. (3) and (5) fitted to the measured capillary pressure curve for sample CT_1. Eq. (3) has seven (i.e., $D_1$, $D_2$, $\beta_1$, $\beta_2$, $P_{d1}$, $P_x$, and $\phi_2$), while Eq. (5) six parameters (i.e., $D_1$, $D_2$, $\beta_1$, $\beta_2$, $P_{d1}$, $P_{d2}$) to be determined through the fitting procedure. Recall that following the proposed approach in this study $P_x$ and $\phi_2$ (or $\phi_1$) in Eq. (5) are calculated from the derived pore size distribution, as shown in Fig. 3b. The optimized parameters for each model are given in Fig. 4 for sample CT_1. As can be seen, both capillary pressure curve models, Eqs. (3) and (5), fit the measured data well. Figure 4 also displays the water relative permeability estimated via Eq. (4) from the optimized parameters of Eqs. (3) and (5) as well as the measured $k_{rw}$ for sample CT_1. The calculated RMSLE values indicate that representing the capillary pressure curve using Eq. (5) results in more accurate estimations of $k_{rw}$ than Eq. (3). Similar results were obtained for other samples (not shown). Figure 5 demonstrates the measured $k_{rw}$ as well as the estimated $k_{rw}$ using the bi-modal CPA model, Eq. (4), and the bi-modal van Genuchten-Mualem model, Eq. (2), for CT samples. The RMSLE value calculated for each model is also reported. Results show that the CPA model estimated $k_{rw}$ more accurately than the vG-M model (see RMSLE values). We found that the bi-modal CPA model estimated water relative permeability over the entire range of water saturation satisfactorily. Although the bi-modal vG-M model estimated $k_{rw}$ at high water saturations ($0.8 < S_w < 1$), it mainly overestimated water relative permeability at intermediate and low saturations ($S_w < 0.8$). Similar results were obtained by Priesack and Durner (2006) and Spohrer et al. (2006). For example, Priesack and Durner (2006) argued that adapting the exponent 3 instead of 2 in Eq. (2) resulted into more accurate $k_{rw}$ estimation for a microporous silt loam.

As can be seen in Fig. 5, there is a sharp change in slope at the crossover saturation $S_{wx}$ on the $k_{rw}$ curves estimated via CPA, in contrast to the smooth curves estimated by the series-parallel tubes approach. Hunt et al. (2013) also demonstrated that the estimated $k_{rw}$ via the bi-modal CPA model sharply switches from the structure domain to the texture one (see their Fig. 4). As those authors discussed, the CPA model is sensitive to the value of fractal dimension, which mainly controls the shape and slope of the $k_{rw}$ curve. Therefore, even a small difference between fractal dimensions of the structure and texture domains may cause a sharp change in slope at $S_{wx}$. In other words, the less the discrepancy between $D_1$ and $D_2$, the smoother the $k_{rw}$ curve.

Figures 6 displays the measured and estimated $k_{rw}$ curves for NT samples. Similar to the results for CT samples presented in Fig. 5, the bi-modal CPA model, Eq. (4), estimated $k_{rw}$ more precisely than the vG-M model, Eq. (2), for NT samples. This can be interpreted from the RMSLE values given in Fig. 6. By comparing Figs. (5) and (6), one may notice that the bi-modal vG-M model estimated water relative permeability for NT samples better than CT samples. We found that the average RMSLE value for the bi-modal CPA and vG-M models were 0.47 and 2.95, respectively, for CT samples. Interestingly, the average RMSLE value for the bi-modal CPA model increased to 0.87, while for the bi-modal vG-M model decreased to 2.08 for NT samples. Nonetheless, the accuracy of the bi-modal CPA model is still higher than the bi-modal vG-M model in NT samples (0.87 vs. 2.08).

We should point out that the CPA model estimates $k_{rw}$ from the measured saturated volumetric water content $\theta_s$ and five optimized parameters i.e., $D_1$, $D_2$, $\beta_1$, $\beta_2$ and $\phi_2$. However, the vG-M approach has five parameters i.e., $w_1$, $\alpha_1$, $\alpha_2$, $m_1$ and $m_2$; one less than CPA. Given that $\theta_s$ is one measured point on the capillary pressure curve, both models estimate water relative permeability from the same measured $P_c$-$S_w$ curve. Comparing RMSLE values given in Figs. (5) and (6) indicates that the RMSLE values for CPA are substantially less than those for vG-M. Such discrepancies between RMSLE values can not be attributed to incorporating $\theta_s$ measurement within the CPA approach. Instead, we believe that critical path analysis from statistical physics provides a more powerful and realistic platform for the estimation of water relative permeability from capillary pressure curve. Our results are in accord with those reported by Ghanbarian-Alavijeh and Hunt (2012) who demonstrated that CPA estimated $k_{rw}$ from $P_c$-$S_w$ curve for 104 soil samples with uni-modal pore size distributions better than vG-M.

## 4. Conclusions

In this study, we modified the bi-modal capillary pressure curve model proposed by Hunt et al. (2013) to better represent the discontinuity in the power-law capillary pressure curve model, detect $S_{wx}$ and $P_x$ values, and more accurately estimate water relative permeability. To estimate $k_{rw}$ from the $P_c$-$S_w$ curve, we invoked two techniques: (1) critical path analysis, and (2) series-parallel tubes. We used two datasets each of which consisted of 13 soil samples. In the first dataset, the cultivation condition was conventional tillage (CT), while in the second dataset there was no tillage (NT). In both datasets, capillary pressure and water relative permeability curves were measured accurately at 500 water saturations. Results showed that the bi-modal CPA model estimated $k_{rw}$ more accurately than the bi-modal vG-M model for CT and NT samples. We also found that accurate estimation of $k_{rw}$ via CPA requires precise characterization of capillary pressure curve as well as the crossover point separating the structure domain from the texture one.

**Acknowledgment**

The author acknowledges Andreas Schwen, Austrian Agency for Health and Food Safety Vienna Austria, for sharing experimental data used in this study and Yashar Mehmani, Department of Energy Resources Engineering Stanford University, for his fruitful comments. The Author is also grateful to Kansas State University for supports through faculty startup funds.

properties. Soil Sci. 158, 77–85.

Table 1. Properties of soil samples collected under conventional tillage (CT) and non-tillage (NT) conditions.

| Sample | Schwen et al. (2015) | | | | | | | This study | | | | | | | |
|---|---|---|---|---|---|---|---|---|---|---|---|---|---|---|---|
| | $\phi$ | $\theta_s$ | $\alpha_1$ (m$^{-1}$) | $n_1^*$ | $w_2$ | $\alpha_2$ (m$^{-1}$) | $n_2$ | $D_1$ | $\beta_1$ | $P_{d1}$ (cm) | $D_2$ | $\beta_2$ | $P_{d2}$ (cm) | $P_x$ (cm) | $\phi_1$ | $\phi_2$ |
| CT_1 | 0.50 | 0.47 | 30.20 | 1.226 | 0.26 | 0.03 | 1.930 | 2.875 | 0.47 | 1.8 | 2.571 | 0.22 | 1418 | 679 | 0.25 | 0.22 |
| CT_2 | 0.49 | 0.46 | 44.67 | 1.192 | 0.21 | 0.01 | 3.394 | 2.871 | 0.44 | 1.3 | 2.511 | 0.17 | 3508 | 2590 | 0.27 | 0.19 |
| CT_3 | 0.56 | 0.50 | 9.87 | 2.174 | 0.43 | 0.06 | 1.401 | 2.235 | 0.30 | 6.6 | 2.687 | 0.22 | 1105 | 373 | 0.28 | 0.22 |
| CT_4 | 0.56 | 0.49 | 16.27 | 1.708 | 0.42 | 0.04 | 1.437 | 2.511 | 0.31 | 3.9 | 2.637 | 0.22 | 1696 | 469 | 0.28 | 0.22 |
| CT_5 | 0.51 | 0.46 | 13.10 | 1.754 | 0.53 | 0.06 | 1.414 | 2.489 | 0.24 | 5.0 | 2.601 | 0.24 | 1527 | 373 | 0.22 | 0.24 |
| CT_6 | 0.49 | 0.45 | 16.04 | 1.611 | 0.52 | 0.05 | 1.419 | 2.573 | 0.24 | 4.2 | 2.606 | 0.25 | 1423 | 340 | 0.20 | 0.25 |
| CT_7 | 0.55 | 0.50 | 10.17 | 2.131 | 0.43 | 0.05 | 1.420 | 2.194 | 0.30 | 6.8 | 2.636 | 0.22 | 1564 | 373 | 0.29 | 0.22 |
| CT_8 | 0.55 | 0.50 | 25.39 | 1.656 | 0.45 | 0.07 | 1.373 | 2.535 | 0.30 | 2.6 | 2.631 | 0.23 | 1245 | 310 | 0.27 | 0.23 |
| CT_9 | 0.56 | 0.48 | 31.59 | 1.584 | 0.41 | 0.08 | 1.364 | 2.595 | 0.31 | 2.0 | 2.638 | 0.21 | 1048 | 296 | 0.27 | 0.21 |
| CT_10 | 0.60 | 0.48 | 9.97 | 2.086 | 0.46 | 0.22 | 1.313 | 2.347 | 0.30 | 6.6 | 2.690 | 0.22 | 501.1 | 224 | 0.26 | 0.22 |
| CT_11 | 0.53 | 0.47 | 12.69 | 1.239 | 0.16 | 0.01 | 3.378 | 2.852 | 0.49 | 4.5 | 2.511 | 0.16 | 2863 | 2360 | 0.29 | 0.18 |
| CT_12 | 0.54 | 0.41 | 42.65 | 1.238 | 0.40 | 0.07 | 1.460 | 2.935 | 0.58 | 1.1 | 2.620 | 0.25 | 835.6 | 195 | 0.16 | 0.25 |
| CT_13 | 0.54 | 0.50 | 21.78 | 1.636 | 0.41 | 0.06 | 1.363 | 2.551 | 0.32 | 3.0 | 2.645 | 0.22 | 1340 | 340 | 0.28 | 0.22 |
| NT_1 | 0.40 | 0.40 | 2.99 | 1.179 | 0.35 | 0.02 | 3.324 | 2.958 | 0.74 | 16.0 | 2.447 | 0.25 | 2293 | 1080 | 0.12 | 0.28 |
| NT_2 | 0.45 | 0.45 | 3.83 | 1.194 | 0.37 | 0.03 | 2.765 | 2.945 | 0.70 | 13.9 | 2.408 | 0.29 | 1730 | 816 | 0.14 | 0.31 |
| NT_3 | 0.41 | 0.39 | 1.87 | 1.172 | 0.48 | 0.03 | 2.651 | 2.928 | 0.34 | 26.2 | 2.561 | 0.30 | 904.8 | 539 | 0.07 | 0.32 |
| NT_4 | 0.47 | 0.42 | 20.00 | 1.153 | 0.39 | 0.02 | 2.786 | 2.970 | 0.86 | 2.3 | 2.550 | 0.25 | 1106 | 679 | 0.14 | 0.28 |
| NT_5 | 0.49 | 0.45 | 6.89 | 1.603 | 0.64 | 0.04 | 1.454 | 2.662 | 0.20 | 8.8 | 2.646 | 0.31 | 1412 | 409 | 0.15 | 0.30 |
| NT_6 | 0.46 | 0.46 | 2.82 | 1.211 | 0.29 | 0.02 | 5.901 | 2.951 | 0.91 | 16.3 | 2.458 | 0.26 | 1575 | 1420 | 0.18 | 0.28 |
| NT_7 | 0.44 | 0.41 | 13.02 | 1.161 | 0.35 | 0.02 | 3.258 | 2.951 | 0.61 | 3.8 | 2.515 | 0.24 | 1556 | 1080 | 0.15 | 0.26 |
| NT_8 | 0.44 | 0.44 | 2.18 | 1.194 | 0.31 | 0.02 | 6.054 | 2.962 | 1.00 | 21.3 | 2.515 | 0.25 | 2050 | 2050 | 0.16 | 0.28 |
| NT_9 | 0.45 | 0.45 | 4.61 | 1.198 | 0.29 | 0.02 | 3.403 | 2.959 | 1.00 | 9.8 | 2.526 | 0.25 | 1677 | 1290 | 0.18 | 0.27 |
| NT_10 | 0.45 | 0.44 | 13.89 | 1.184 | 0.51 | 0.03 | 1.694 | 2.974 | 1.00 | 3.5 | 2.512 | 0.32 | 1734 | 282 | 0.11 | 0.33 |
| NT_11 | 0.41 | 0.40 | 4.33 | 1.183 | 0.32 | 0.02 | 3.995 | 2.968 | 1.00 | 10.2 | 2.518 | 0.23 | 1964 | 1490 | 0.15 | 0.26 |
| NT_12 | 0.42 | 0.42 | 4.51 | 1.184 | 0.30 | 0.02 | 7.129 | 2.965 | 1.00 | 9.7 | 2.545 | 0.24 | 1614 | 1560 | 0.16 | 0.26 |
| NT_13 | 0.46 | 0.39 | 1.26 | 3.274 | 0.72 | 0.06 | 1.410 | 1.875 | 0.13 | 55.5 | 2.605 | 0.27 | 1743 | 428 | 0.12 | 0.27 |

$^*$ $m_1 = 1 - 1/n_1$ and $m_2 = 1 - 1/n_2$

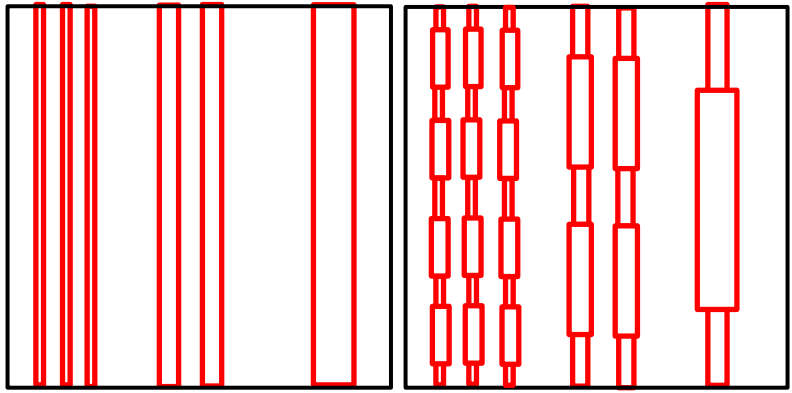

Figure 1. Two-dimensional schematic illustration of the parallel (left) and series-parallel (right) tubes approaches. In the former, pores are replaced with non-interconnected parallel tubes of various sizes. In the latter, however, the porous medium is replaced with the bundle of non-interconnected parallel tubes each of which is composed of capillaries of two different sizes in series.

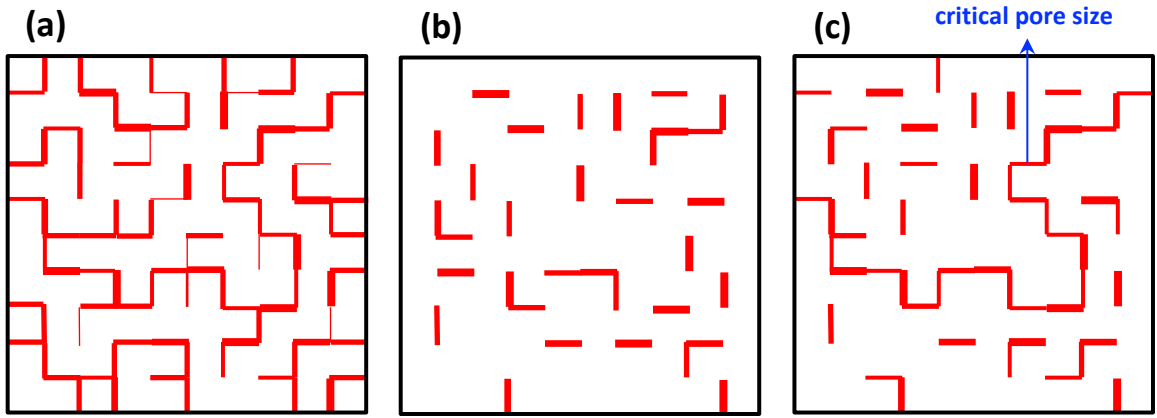

Figure 2. Two-dimensional scheme of the critical path analysis. (a) A pore network compsed of six different pore sizes (i.e., 0.5, 1, 1.5, 2, 2.5 and 3 with arbitrary units) randomly distributed in the medium. (b) The same network with only the first two largest pores (2.5 and 3) in their original locations. Pores smaller than 2.5 were removed from the pore network. As can be seen, the medium does not percolate. (c) The nerwork after adding the third largest pores with size 2 (critical pore size). The sample-spanning cluster is first formed and the network starts percolating.

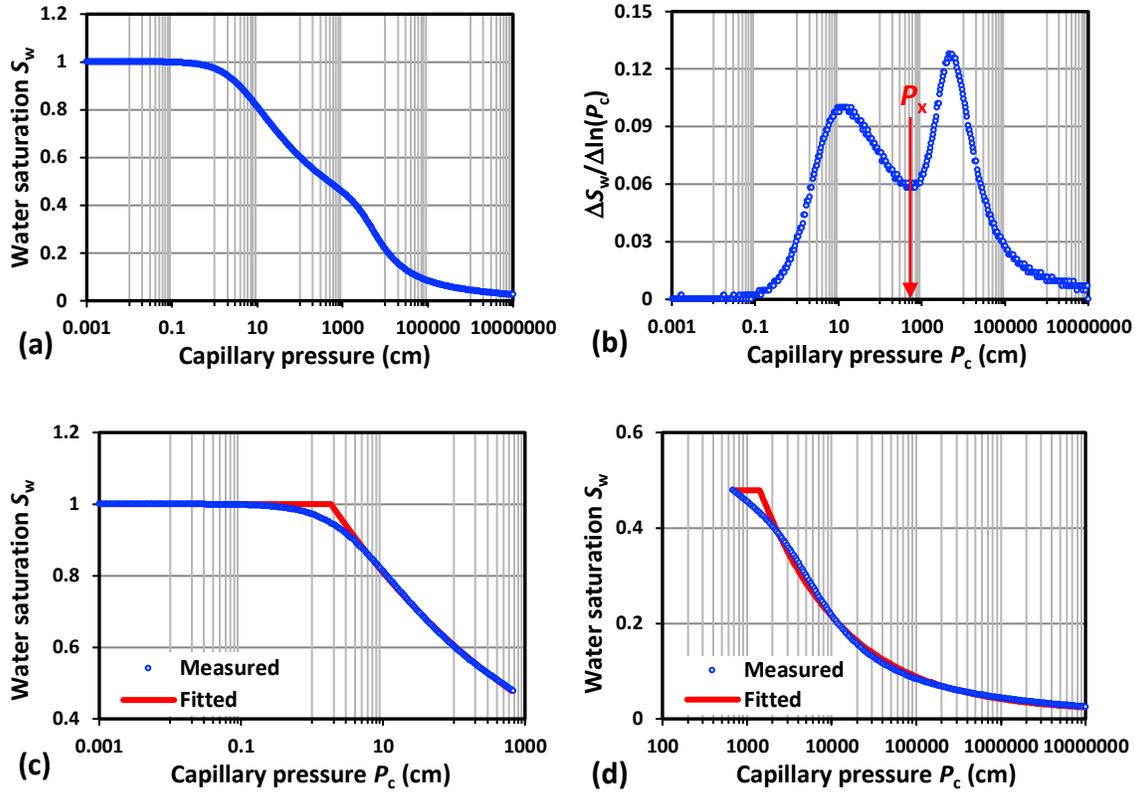

Figure 3. (a) Measured capillary pressure curve, (b) derived bi-modal pore size distribution, (c) fitted Eq. (5) to the capillary pressure data at higher water saturations, and (d) fitted Eq. (5) to the capillary pressure data at lower water saturations for sample CT_1. The optimized parameters of Eq. (5) are given in Table 1.

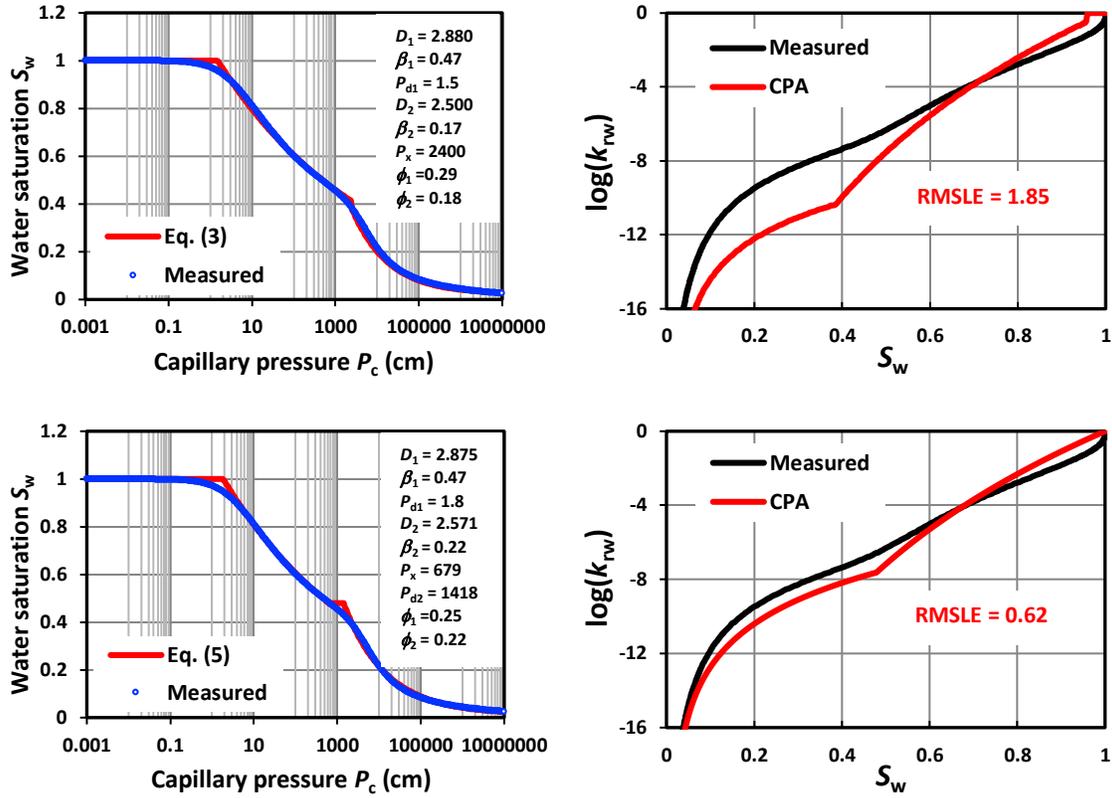

Figure 4. Comparing Eqs. (3) and (5) in the estimation of water relative permeability $k_{rw}$ via the critical path analysis, Eq. (4), for sample CT_1. (left) Measured capillary pressure curves and fitted models, Eq. (3) and (5). (right) Measured and estimated $k_{rw}$ from the fitted capillary pressure curve models and their optimized parameters.

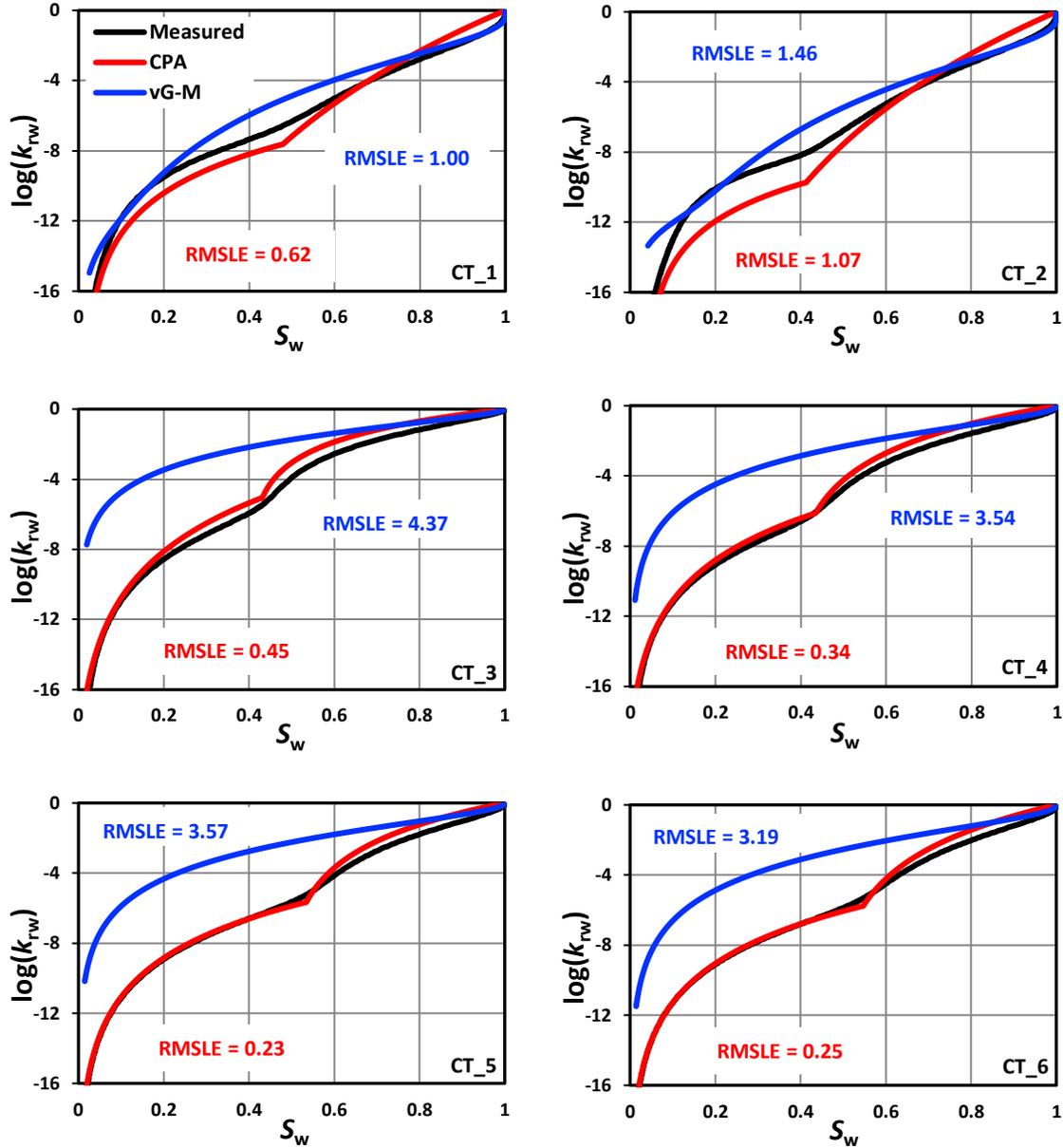

Figure 5. The measured $k_{rw}$ and the estimated $k_{rw}$ using the bi-modal CPA model, Eq. (4) and the bi-modal van Genuchten-Mualem model, Eq. (2), for CT samples. The optimized parameters of the capillary pressure curve models (Eqs. 1 and 5) are given in Table 1.

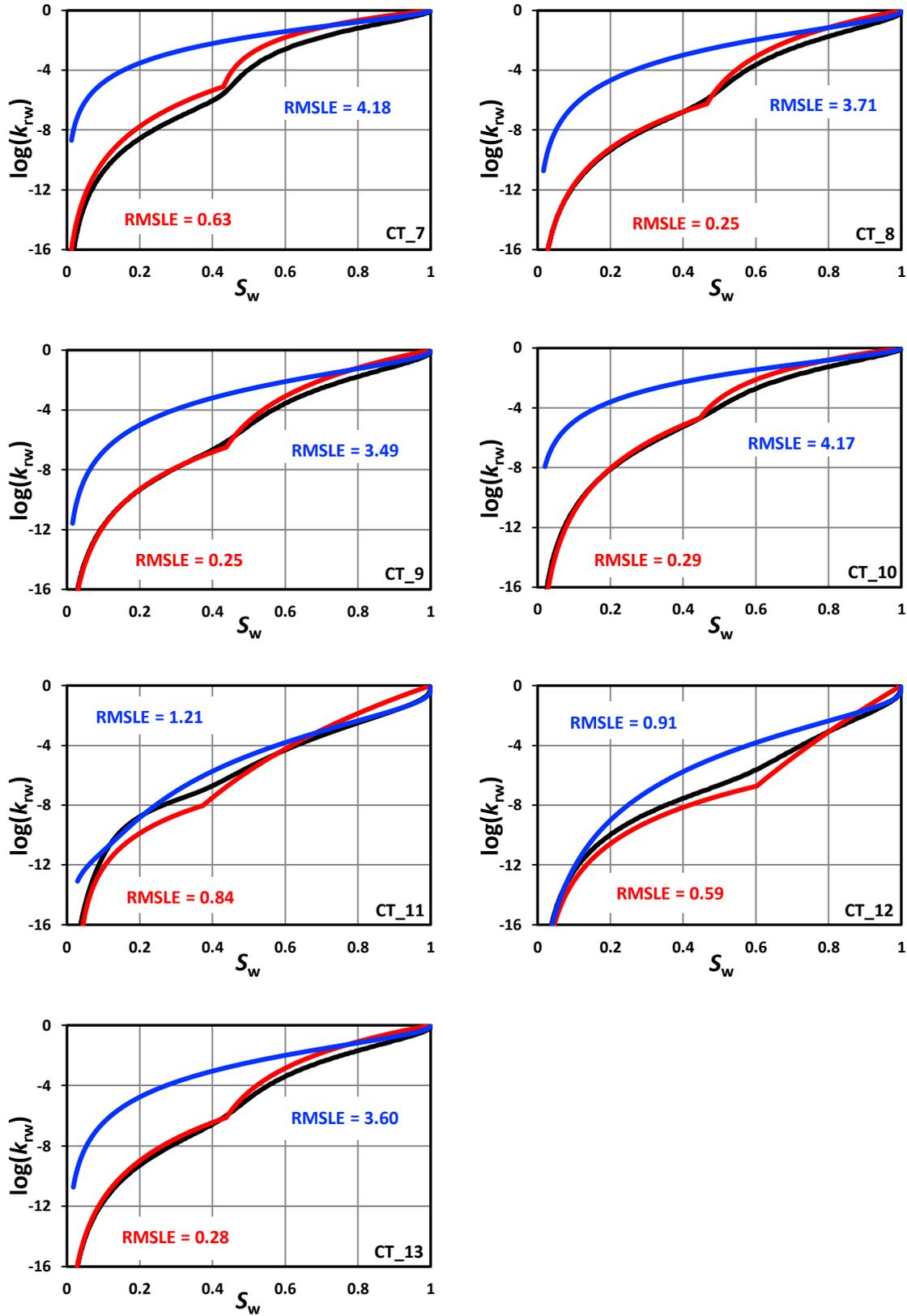

Figure 5. Continued.

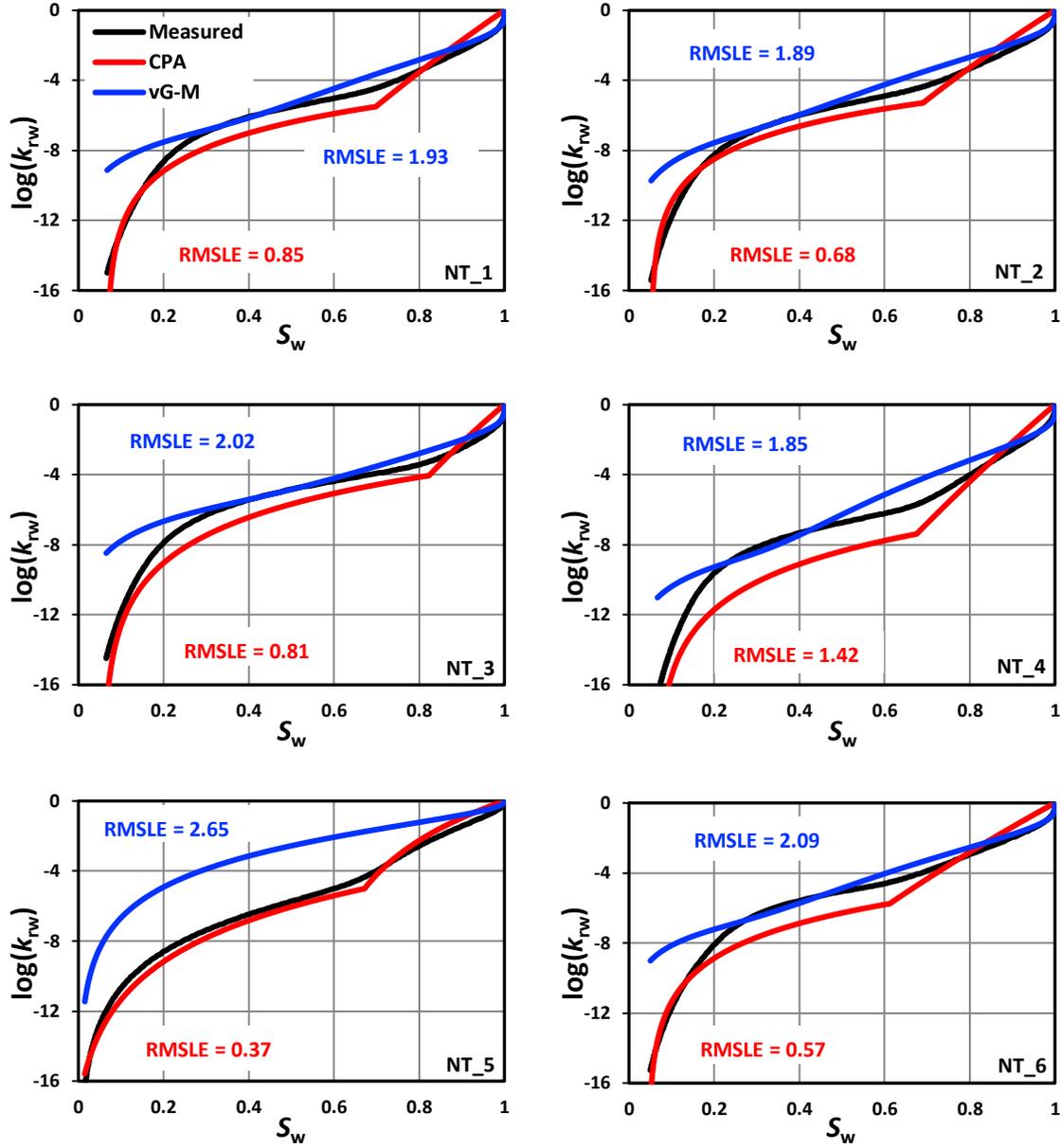

Figure 6. The measured $k_{rw}$ and the estimated $k_{rw}$ using the bi-modal CPA model, Eq. (4) and the bi-modal van Genuchten-Mualem model, Eq. (2), for NT samples. The optimized parameters of the capillary pressure curve models (Eqs. 1 and 5) are given in Table 1.

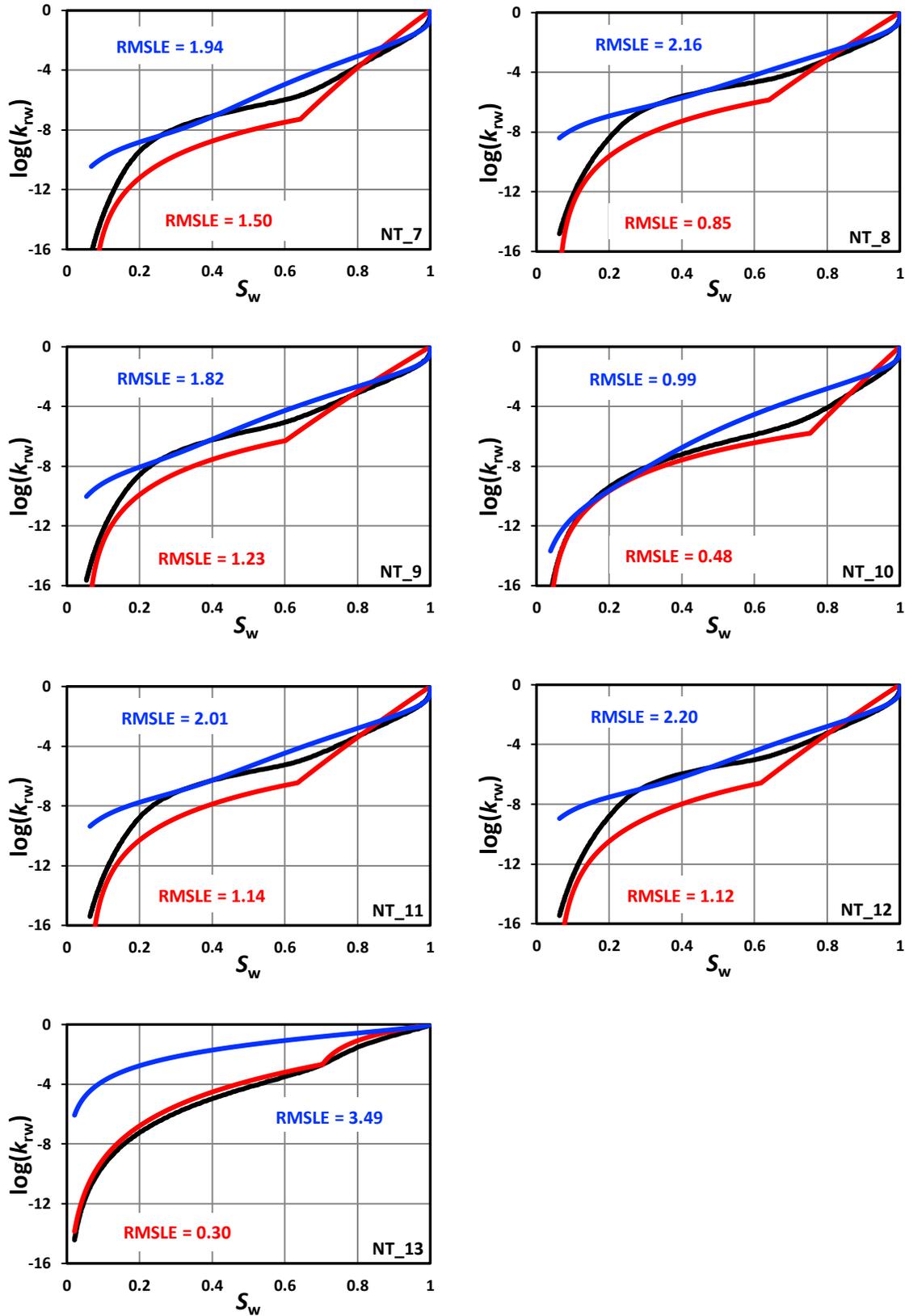

Figure 6. Continued.